\documentclass[10pt]{IEEEtran15}
\usepackage{graphicx}
\usepackage{amsmath}
\usepackage{amssymb}
\usepackage{color}
\usepackage{epsfig}

\newcommand{\Define}{\stackrel{\triangle}{=}}

\begin{document}
\twocolumn

\title{\LARGE Low-Complexity Near-ML Decoding of Large Non-Orthogonal 
STBCs using Reactive Tabu Search}
\author{N. Srinidhi, Saif K. Mohammed, A. Chockalingam, and B. Sundar Rajan \\
\vspace{-0mm}
{\normalsize Department of ECE, Indian Institute of Science,
Bangalore 560012, INDIA} 
\vspace{-8mm}
}
\maketitle
\thispagestyle{empty}

\begin{abstract}
Non-orthogonal space-time block codes (STBC) with {\em large dimensions} 
are attractive because they can simultaneously achieve both high spectral 
efficiencies (same spectral efficiency as in V-BLAST for a given number 
of transmit antennas) {\em as well as} full transmit diversity. Decoding 
of non-orthogonal STBCs with large dimensions has been a challenge. In 
this paper, we present a reactive tabu search (RTS) based algorithm for 
decoding non-orthogonal STBCs from cyclic division algebras (CDA) having 
large dimensions. Under i.i.d fading and perfect channel state information 
at the receiver (CSIR), our simulation results show that RTS based decoding 
of $12\times 12$ STBC from CDA and 4-QAM with 288 real dimensions achieves 
$i)$ $10^{-3}$ uncoded BER at an SNR of just 0.5 dB away from SISO AWGN 
performance, and $ii)$ a coded BER performance close to within about 5 dB 
of the theoretical MIMO capacity, using rate-3/4 turbo code at a spectral 
efficiency of 18 bps/Hz. RTS is shown to achieve near SISO AWGN performance 
with less number of dimensions than with LAS algorithm (which we reported 
recently) at some extra complexity than LAS. We also report good BER 
performance of RTS when i.i.d fading and perfect CSIR assumptions are 
relaxed by considering a spatially correlated MIMO channel model, and 
by using a training based iterative RTS decoding/channel estimation
scheme.
\end{abstract} 

\vspace{-1.0mm}

\vspace{-3.0mm}
\section{Introduction}
\label{sec1}
\vspace{-2.0mm}
MIMO systems that employ non-orthogonal space-time block codes (STBC) 
from cyclic division algebras (CDA) for arbitrary number of transmit 
antennas, $N_t$, are attractive because they can simultaneously provide 
both {\em full-rate} (i.e., $N_t$ complex symbols per channel use, 
which is same as in V-BLAST) as well as {\em full transmit diversity} 
\cite{bsr},\cite{cda}. The $2\times 2$ Golden code is a well known 
non-orthogonal STBC from CDA for 2 transmit antennas \cite{gold05}. 
High spectral efficiencies of the order of tens of bps/Hz can be 
achieved using large non-orthogonal STBCs. For e.g., a $16\times 16$ 
STBC from CDA has 256 complex symbols in it with 512 real dimensions; 
with 16-QAM and rate-3/4 turbo code, this system offers a high spectral 
efficiency of 48 bps/Hz. Decoding of non-orthogonal STBCs with such 
large dimensions, however, has been a challenge. Sphere decoder and its 
low-complexity variants are prohibitively complex for decoding such STBCs 
with hundreds of dimensions. Recently, we proposed a low-complexity near-ML 
achieving algorithm to decode large non-orthogonal STBCs from CDA; this 
algorithm, which is based on bit-flipping approach, is termed as likelihood 
ascent search (LAS) algorithm \cite{jsac}-\cite{gcom08}. In this paper, 
we present a {\em reactive tabu search (RTS) based approach} to near-ML 
decoding of non-orthogonal STBCs with large dimensions. 

Key attractive features of the proposed RTS based decoding are its
low-complexity and near-ML performance in systems with large dimensions
(e.g., hundreds of dimensions). While creating hundreds of dimensions 
in space alone (e.g., V-BLAST) requires hundreds of antennas, use of 
non-ortho\-gonal STBCs from CDA can create hundreds of dimensions with 
just tens of antennas (space) and tens of channel uses (time). Given 
that 802.11 smart WiFi products with 12 transmit antennas\footnote{12 
antennas in these products are now used only for beamforming. Single-beam 
multi-antenna approaches can offer range increase and interference avoidance, 
but not spectral efficiency increase.} at 2.5 GHz are now commercially
available \cite{ruckus} (which establishes that issues related to
placement of many antennas and RF/IF chains can be solved in large
aperture communication terminals like set-top boxes/laptops), large
non-orthogonal STBCs (e.g., $16\hspace{-0.4mm}\times\hspace{-0.4mm} 16$
{\small STBC} from {\small CDA}) in combination with large dimension 
near-ML decoding using RTS can enable communications at increased 
spectral efficiencies of the order of tens of bps/Hz (note that current 
standards achieve only $< 10$ bps/Hz using only up to 4 tx antennas).

Tabu search (TS), a heuristic originally designed to obtain approximate 
solutions to combinatorial optimization problems \cite{tabu1}-\cite{tabu3}, 
is increasingly applied in communication problems \cite{tabu4}-\cite{tabu6}. 
For e.g.,  in \cite{tabu4}, design of constellation label maps to maximize 
asymptotic coding gain is formulated as a quadra\-tic assignment problem 
(QAP), which is solved using RTS \cite{tabu3}. RTS approach is shown to 
be effective in terms of BER performance and efficient in terms of 
computational complexity in CDMA multiuser detection \cite{tabu5}. 
In \cite{tabu6}, a fixed TS based detection in V-BLAST is presented. 
In this paper, we establish that RTS based decoding of non-orthogonal 
STBCs can achieve excellent BER performance (near-ML and near-capacity 
performance) in large dimensions at practically affordable low-complexities. 
We also present a stopping-criteri\-on for the RTS algorithm. RTS for large 
dimension non-orthogonal STBC decoding has not been reported so far. Our 
results in this paper can be summarized as follows:
\begin{itemize}
\vspace{-1.50mm}
\item 	Under i.i.d fading and perfect channel state information
	at the receiver (CSIR), our simulation results show that RTS 
	based decoding of $12\times 12$ STBC from CDA and 4-QAM (288 
	real dimensions) achieves $i)$ $10^{-3}$ uncoded BER at an 
	SNR of just 0.5 dB away from SISO AWGN performance, and $ii)$ 
	a coded BER performance close to within about 5 dB of the 
	theoretical capacity using rate-3/4 turbo code at a spectral 
	efficiency of 18 bps/Hz. 
\item	Compared to the LAS algorithm we reported recently in 
	\cite{jsac}-\cite{gcom08}, RTS achieves near-SISO AWGN 
	performance with less number of dimensions than with LAS;
	this is achieved at some extra complexity compared to LAS.
\item	We report good BER performance when i.i.d fading and perfect 
	CSIR assumptions are relaxed by adopting a spatially correlated 
	MIMO channel model, and a training based iterative RTS
	decoding/channel estimation scheme. 
\end{itemize}

The rest of this paper is organized as follows. The non-ortho\-gonal STBC
MIMO system model is presented in Section \ref{sec2}. RTS algorithm for
decoding non-orthogonal STBCs and the proposed stopping criterion are 
presented in Section \ref{sec3}. Simulation results including uncoded 
and coded BER performance of RTS decoding with $i$) perfect CSIR, 
$ii)$ estimated CSIR using an iterative RTS decoding/channel estimation 
scheme, and $iii)$ effect of spatial correlation are presented in Section
\ref{sec4}. Conclusions are given in Section \ref{sec5}.

\vspace{-3.00mm}
\section{Non-Orthogonal STBC MIMO System Model}
\label{sec2}
\vspace{-0.0mm}
Consider a STBC MIMO system with multiple transmit
and receive antennas. An $(n,p,k)$ STBC is
represented by a matrix {\small ${\bf X}_c \in {\mathbb C}^{n \times p}$},
where $n$ and $p$ denote the number of transmit antennas and number of
time slots, respectively, and $k$ denotes the number of complex data
symbols sent in one STBC matrix. The $(i,j)$th entry in ${\bf X}_c$
represents the complex number transmitted from the $i$th transmit
antenna in the $j$th time slot. The rate of an STBC is $\frac{k}{p}$.
Let $N_r$ and $N_t=n$ denote the number of receive and transmit antennas,
respectively. Let ${\bf H}_c \in {\mathbb C}^{N_r\times N_t}$ denote the
channel gain matrix, where the $(i,j)$th entry in ${\bf H}_c$ is the
complex channel gain from the $j$th transmit antenna to the $i$th receive
antenna. We assume that the channel gains remain constant over one STBC
matrix and vary (i.i.d) from one STBC matrix  to the other. Assuming rich
scattering, we model the entries of ${\bf H}_c$ as i.i.d
$\mathcal C \mathcal N(0,1)$. The received space-time signal matrix,
{\small ${\bf Y}_c \in {\mathbb C}^{N_r \times p}$}, can be written as
\begin{equation}
\label{SystemModel}
{\bf Y}_c = {\bf H}_c{\bf X}_c + {\bf N}_c,
\end{equation}
where ${\bf N}_c \in {\mathbb C}^{N_r \times p}$ is the noise matrix
at the receiver and its entries are modeled as i.i.d
$\mathcal C \mathcal N\big(0,\sigma^2=\frac{N_tE_s}{\gamma}\big)$,
where $E_s$ is the average energy of the transmitted symbols, and
$\gamma$ is the average received SNR per receive antenna \cite{jafarkhani},
and the $(i,j)$th entry in ${\bf Y}_c$ is the received signal at the $i$th
receive antenna in the $j$th time-slot. Consider linear dispersion STBCs,
where ${\bf X}_c$ can be written in the form \cite{jafarkhani}
\begin{eqnarray}
\label{SystemModelStbcLpx}
{\bf X}_c & = & \sum_{i = 1}^{k} x_c^{(i)} {\bf A}_c^{(i)},
\end{eqnarray}
where $x_c^{(i)}$ is the $i$th complex data symbol, and
${\bf A}_c^{(i)} \in {\mathbb C}^{N_t \times p}$ is its
corresponding weight matrix. The received signal model in
(\ref{SystemModel}) can be written in an equivalent V-BLAST
form as
\begin{equation}
\label{SystemModelvec2}
{\bf y}_c \,\, = \,\, \sum_{i=1}^{k} x_c^{(i)}\, (\widehat{{\bf H}}_c\, {\bf a}_c^{(i)}) + {\bf n}_c \,\, = \,\, \widetilde{{\bf H}}_c {\bf x}_c + {\bf n}_c,
\end{equation}
where
${\bf y}_c \in {\mathbb C}^{N_rp \times 1} = vec\,({\bf Y}_c)$,
$\widehat{{\bf H}}_c \in {\mathbb C}^{N_rp \times N_tp} = ({\bf I} \otimes {\bf H}_c)$,
${\bf a}_c^{(i)} \in {\mathbb C}^{N_tp \times 1} = vec\,({\bf A}_c^{(i)})$,
${\bf n}_c \in {\mathbb C}^{N_rp \times 1} = vec\,({\bf N}_c)$,
${\bf x}_c \in {\mathbb C}^{k \times 1}$
whose $i$th entry is the data symbol $x_c^{(i)}$, and
$\widetilde{{\bf H}}_c \in {\mathbb C}^{N_rp \times k}$ whose
$i$th column is
$\widehat{{\bf H}}_c \, {\bf a}_c^{(i)}$, $i=1,2,\cdots,k$.
Each element of ${\bf x}_c$ is an $M$-PAM/$M$-QAM symbol.
Let ${\bf y}_c$, $\widetilde{{\bf H}}_c$, ${\bf x}_c$, ${\bf n}_c$
be decomposed into real and imaginary parts as: 
\begin{eqnarray}
\label{SystemModelDecompose}
{\bf y}_c={\bf y}_I + j{\bf y}_Q, &  {\bf x}_c = {\bf x}_I + j{\bf x}_Q,
\nonumber \\
{\bf n}_c={\bf n}_I+j{\bf n}_Q, & \widetilde{{\bf H}}_c={\bf H}_I+j{\bf H}_Q.
\end{eqnarray}
Further, we define
${\bf H}_r \in {\mathbb R}^{2N_rp \times 2k}$,
${\bf y}_r \in {\mathbb R}^{2N_rp \times 1}$,
${\bf x}_r \in {\mathbb R}^{2k \times 1}$, and
${\bf n}_r \in {\mathbb R}^{2N_rp \times 1}$ as
\begin{eqnarray}
\label{SystemModelRealDef}
{\bf H}_r = \left(\begin{array}{cc}{\bf H}_I \hspace{2mm} -{\bf H}_Q \\
{\bf H}_Q  \hspace{5mm} {\bf H}_I \end{array}\right),
\hspace{4mm}
{\bf y}_r = [{\bf y}_I^T \hspace{2mm} {\bf y}_Q^T ]^T, \\
\hspace{4mm}
{\bf x}_r = [{\bf x}_I^T \hspace{2mm} {\bf x}_Q^T ]^T,
\hspace{4mm}
{\bf n}_r = [{\bf n}_I^T \hspace{2mm} {\bf n}_Q^T ]^T.
\end{eqnarray}
Now, (\ref{SystemModelvec2}) can be written as
\vspace{-2mm}
\begin{eqnarray}
\label{SystemModelReal}
{\bf y}_r & = & {\bf H}_r{\bf x}_r + {\bf n}_r.
\end{eqnarray}

\vspace{-4mm}
Henceforth, we work with the real-valued system in
(\ref{SystemModelReal}). For notational simplicity, we drop
subscripts $r$ in (\ref{SystemModelReal}) and write
\begin{eqnarray}
\label{SystemModelII}
{\bf y} & = & {\bf H} {\bf x} + {\bf n},
\end{eqnarray}
where {\small ${\bf H} = {\bf H}_r \in {\mathbb R}^{2N_rp \times 2k}$,
${\bf y} = {\bf y}_r \in {\mathbb R}^{2N_rp \times 1}$,
${\bf x} = {\bf x}_r \in {\mathbb R}^{2k \times 1}$}, and
{\small ${\bf n} = {\bf n}_r \in {\mathbb R}^{2N_rp \times 1}$.}
We assume that the channel coefficients are known at the receiver
but not at the transmitter. Let 
${\mathbb A} \Define \{a_q, q=1,2,\cdots,M\},$ where $a_q=2q-1-M$ 
denote the $M$-PAM signal set from which $x_i$ ($i$th entry of 
${\bf x}$) takes values, $i=0,\cdots,2k-1$. The ML solution is given by
\begin{eqnarray}
\label{MLdetection}
{\bf d}_{ML} & = & {\mbox{arg min}\atop{{\bf d} \in {\mathbb A}^{2k}}}
\thinspace
{\bf d}^T {\bf H}^T{\bf H}{\bf d} - 2{\bf y}^T{\bf H}{\bf d},
\end{eqnarray}
whose complexity is exponential in $k$.

\vspace{-4.0mm}
\subsection{Full-rate Non-orthogonal STBCs from CDA}
\vspace{-1.0mm}
We focus on the decoding of square (i.e., 
$n\hspace{-0.9mm}=\hspace{-0.9mm}p\hspace{-0.9mm}=\hspace{-0.9mm}N_t$),
full-rate (i.e., 
$k\hspace{-0.9mm}=\hspace{-0.9mm}pn\hspace{-0.9mm}=\hspace{-0.9mm}N_t^2$),
circulant (where the weight matrices ${\bf A}_c^{(i)}$'s are permutation
type), non-orthogonal STBCs from CDA \cite{bsr}, whose construction for
arbitrary number of transmit antennas $n$ is given by the matrix in
Eqn.(9.a) given at the bottom of the next page. In (9.a),
{\small $\omega_n=e^{\frac{{\bf j}2\pi}{n}}$, ${\bf j}=\sqrt{-1}$,
and $d_{u,v}$, $0\leq u,v \leq n-1$} are the $n^2$ data symbols from 
a QAM alphabet. When $\delta=t=1$, the code in (9.a) is information 
lossless (ILL), and when $\delta=e^{\sqrt{5}\,{\bf j}}$ and
$t=e^{{\bf j}}$, it is of full-diversity and information lossless 
(FD-ILL) \cite{bsr}. High spectral efficiencies with large $n$ can be 
achieved using this code construction. However, since these STBCs are 
non-orthogonal, ML detection gets increasingly impractical for large $n$. 
Consequently, a key challenge in realizing the benefits of these large 
STBCs in practice is that of achieving near-ML performance for large $n$ 
at low decoding complexities. The BER performance results we report in
Sec. \ref{sec4} show that the RTS based decoding algorithm we present
in the following section essentially meets this challenge.

\vspace{-3mm}
\section{RTS Algorithm for Large Non-Orthogonal STBC Decoding}
\label{sec3}
\vspace{-1mm}
In this section, we present the RTS algorithm, which is an iterative 
local search algorithm, for decoding non-orthogonal {\small STBC}s. The goal 
is to get $\widehat{{\bf x}}$, an estimate of ${\bf x}$, given 
${\bf y}$ and ${\bf H}$. 

{\em Neighborhood Definitions:} For each vector in the solution space, 
define the neighborhood structure as follows. {\em Symbol neighborhood} 
of a signal point $a_q \in {\mathbb A}$, {\small $q=1,2,\cdots,M$}, is 
defined as a set ${\cal N}(a_q) \subset {\mathbb A} - \{a_q\}$; e.g., 
for 4-PAM, {\small ${\mathbb A}=\{-3,-1,1,3\}$}, one possible symbol 
neighborhood structure could be {\small ${\cal N}(-3)=\{-1,1\}$},
{\small ${\cal N}(-1)=\{-3,1\}$}, {\small ${\cal N}(1)=\{-1,3\}$},
{\small ${\cal N}(3)=\{1,-1\}$}. Then, 
$N\Define |{\cal N}(a_q)|, \forall q\in \{1 \cdots M\}$ is the number 
of symbol neighbors of $a_q$. Note that the maximum and minimum value 
$N$ can take is $M-1$ and 1, respectively. Let 
${\bf x}^{(m)}=\small{[x_0^{(m)} \thinspace x_1^{(m)} \cdots x_{2k-1}^{(m)}]}$
denote the data vector in the $m$th iteration. We refer to the vector 
\begin{eqnarray}
\mathbf{z}^{(m)}(u,v) \,= \, \big[z^{(m)}_0(u,v) 
\,\,\,\, z^{(m)}_1(u,v) \, \cdots \,
z^{(m)}_{2k-1}(u,v) \big], 
\end{eqnarray}
as the $(u,v)$th {\em vector neighbor} of $\mathbf{x}^{(m)}$, 
{\small $u=0,\cdots,2k-1$, $v=0,\cdots,N-1$}, if $1)$ $\mathbf{x}^{(m)}$ 
differs from $\mathbf{z}^{(m)}(u,v)$ in the $u$th coordinate, and $2)$ the 
$u$th element of $\mathbf{z}^{(m)}(u,v)$ is the $v$th symbol neighbor of
$x_u^{(m)}$. That is,
\begin{equation}
z_i^{(m)}(u,v) \,\,= \,\, \left\{ 
\begin{array}{ll}
x^{(m)}_i & \mbox{for} \,\,\, i\neq u \\
w_v(x_u^{(m)}) & \mbox{for} \,\,\, i=u,
\end{array}\right.
\label{eq11}
\end{equation}
where $w_v(a)$, $v=0,1,\cdots,N-1$ is the $v$th element in ${\cal N}(a)$.
So we will have $2kN$ vectors which differ from a given vector 
in the solution space in only one coordinate. These $2kN$ vectors form the
neighborhood of the given vector. It is noted that bit-flipping 
is a special case with $N=1$ and $M=2$.

The algorithm is said to execute a move $(u,v)$ if
$\mathbf{x}^{(m+1)} = \mathbf{z}^{(m)}(u,v)$.
The number of candidates to be considered for a move in the $m$th 
iteration is $2kN$. Since the coordinate that changes in a move can 
take $M$ possible values for $M$-PAM, the total number of possible 
moves is $2kMN$. The tabu value of a move, which is a non-negative 
integer, means that the move cannot be considered for that many number 
of subsequent iterations, unless certain conditions are satisfied.

{\em Tabu Matrix:}
A $\textit{tabu\_matrix}$ of size $2kM\times N$ is the matrix whose 
entries denote the tabu values of moves. The $(r,s)$th entry of the 
$\textit{tabu\_matrix}$ corresponds to the move $(u,v)$ from 
$\mathbf{x}^{(m)}$ when $u=\lfloor \frac{r-1}{M} \rfloor$, $v=s$ 
and $x_u^{(m)}=a_q$, where $q=mod(r-1,M)+1$.

{\em RTS Algorithm:}
Let $\mathbf{g}^{(m)}$ be the vector which has the least ML cost found 
till the $m$th iteration of the algorithm. Let $l_{rep}$ be the average 
length (in number of iterations) between two successive occurrences of
the same solution vector (repetitions), at the end of an iteration. 
Tabu period, $P$, a dynamic non-negative integer parameter, is defined.
If a move is marked as tabu in an iteration, it will remain as tabu 
for $P$ subsequent iterations. The algorithm starts with an initial 
solution vector  ${\bf x}^{(0)}$, which, for e.g., could be the 
MMSE or MF output vector. Set $\mathbf{g}^{(0)} = {\bf x}^{(0)}$, 
$l_{rep}=0$, and $P=P_0$. All the entries of the $\textit{tabu\_matrix}$
are set to zero. The following steps 1) to 3) are performed in each 
iteration. Consider $m$th iteration in the algorithm, $m\geq 0$. 

\thanks{
{\small
\line(1,0){505}
\[
\label{eqn}
\hspace{1.3cm}
\left[
\begin{array}{ccccc}
\sum_{i=0}^{n-1}d_{0,i}\,t^i & \delta\sum_{i=0}^{n-1}d_{n-1,i}\,\omega_n^i\,t^i & \delta\sum_{i=0}^{n-1}d_{n-2,i}\,\omega_n^{2i}\,t^i & \cdots & \delta\sum_{i=0}^{n-1}d_{1,i}\,\omega_n^{(n-1)i}\,t^i \\
\sum_{i=0}^{n-1}d_{1,i}\,t^i & \sum_{i=0}^{n-1}d_{0,i}\,\omega_n^i\,t^i & \delta\sum_{i=0}^{n-1}d_{n-1,i}\,\omega_n^{2i}\,t^i & \cdots & \delta\sum_{i=0}^{n-1}d_{2,i}\,\omega_n^{(n-1)i}\,t^i \\
\sum_{i=0}^{n-1}d_{2,i}\,t^i & \sum_{i=0}^{n-1}d_{1,i}\,\omega_n^i\,t^i & \sum_{i=0}^{n-1}d_{0,i}\,\omega_n^{2i}\,t^i & \cdots & \delta\sum_{i=0}^{n-1}d_{3,i}\,\omega_n^{(n-1)i}\,t^i \\
\vdots & \vdots & \vdots & \vdots & \vdots \\
\sum_{i=0}^{n-1}d_{n-2,i}\,t^i & \sum_{i=0}^{n-1}d_{n-3,i}\,\omega_n^i\,t^i & \sum_{i=0}^{n-1}d_{n-4,i}\,\omega_n^{2i}\,t^i & \cdots & \delta \sum_{i=0}^{n-1}d_{n-1,i}\,\omega_n^{(n-1)i}t^i \\
\sum_{i=0}^{n-1}d_{n-1,i}\,t^i & \sum_{i=0}^{n-1}d_{n-2,i}\,\omega_n^i\,t^i & \sum_{i=0}^{n-1}d_{n-3,i}\,\omega_n^{2i}\,t^i & \cdots & \sum_{i=0}^{n-1}d_{0,i}\,\omega_n^{(n-1)i}\,t^i
\end{array}
\right]. \hspace{10mm} (\mbox{9.a})
\]
}
}

\newpage 
{\em Step 1):} 
Define $\thinspace \mathbf{y}_{mf} \Define \mathbf{H}^T\mathbf{y}$,  
$\thinspace\mathbf{R} \Define \mathbf{H}^T\mathbf{H}$, and
$\thinspace\mathbf{f}^{(m)}\Define \mathbf{R}\mathbf{x}^{(m)}-\mathbf{y}_{mf}$. 
Let $\mathbf{e}^{(m)}(u,v)=\mathbf{z}^{(m)}(u,v)-\mathbf{x}^{(m)}$.
The ML costs of the $2kN$ neighbors of $\mathbf{x}^{(m)}$, namely,
$\mathbf{z}^{(m)}(u,v)$, $u=0,\cdots,2k-1$; $v=0,\cdots,N-1$, are 
computed as

\vspace{-6mm}
{\small 
\begin{eqnarray}
\phi(\mathbf{z}^{(m)}(u,v)) &\hspace{-2mm}=& \hspace{-2mm} \big(\mathbf{x}^{(m)} + \mathbf{e}^{(m)}(u,v)\big)^T \mathbf{R} \thinspace \big(\mathbf{x}^{(m)}+\mathbf{e}^{(m)}(u,v)\big) \nonumber \\
& \hspace{-2mm} & \hspace{-2mm} - 2\big(\mathbf{x}^{(m)}+\mathbf{e}^{(m)}(u,v)\big)^T \mathbf{y}_{mf} \nonumber \\
&\hspace{-35mm}=& \hspace{-18mm} \phi(\mathbf{x}^{(m)})+2\big(\mathbf{e}^{(m)}(u,v)\big)^T\mathbf{R}\thinspace\mathbf{x}^{(m)} \nonumber \\
& \hspace{-18mm} & \hspace{-18mm} +\thinspace \big(\mathbf{e}^{(m)}(u,v)\big)^T\mathbf{R}\thinspace \mathbf{e}^{(m)}(u,v) - 2\big(\mathbf{e}^{(m)}(u,v)\big)^T\mathbf{y}_{mf} \nonumber \\
&\hspace{-35mm}=& \hspace{-18mm} \phi(\mathbf{x}^{(m)})+ \underbrace{2 \thinspace e_u^{(m)}(u,v) \thinspace {f}_u^{(m)} + \thinspace \big(e_u^{(m)}(u,v)\big)^2 \thinspace \mathbf{R}_{u,u}}_{\Define \thinspace C\big(e_u^{(m)}(u,v)\big)}\, ,
\label{cost}
\end{eqnarray}
}

\vspace{-5mm}
where $e_u^{(m)}(u,v)$ is the $u$th element of $\mathbf{e}^{(m)}(u,v)$, 
$f_u^{(m)}$ is $u$th element of $\mathbf{f}^{(m)}$, and
$\mathbf{R}_{u,u}$ is the $(u,u)$th element of $\mathbf{R}$.
$\phi(\mathbf{x}^{(m)})$ on the RHS in (\ref{cost}) can be dropped 
since it will not affect the cost minimization. Let 
\begin{eqnarray}
\hspace{-7mm}
(u_1,v_1)&\hspace{-1mm}=&\hspace{-1mm} 
{\mbox{arg min}\atop{u,v}} \thinspace \thinspace \thinspace
C\big(e_u^{(m)}(u,v)\big).
\end{eqnarray}

\vspace{-3mm}
The move $(u_1,v_1)$ is accepted if any one of the following two
conditions is satisfied:

$i)$ $\phi(\mathbf{z}^{(m)}(u_1,v_1)) < \phi(\mathbf{g}^{(m)})$

$ii)$ {\small $\textit{tabu\_matrix}((u_1-1)M+q,v_1)=0$}
where {\small $q: x_{u_1}^{(m)}=a_q \in \mathbb{A}$}. 

If move $(u_1,v_1)$ is accepted, then make 
\vspace{-1mm}
\begin{eqnarray}
\mathbf{x}^{(m+1)} & = & \mathbf{x}^{(m)}+\mathbf{e}^{(m)}(u_1,v_1).
\end{eqnarray} 

\vspace{-3mm}
If move $(u_1,v_1)$ is not accepted (i.e., neither of conditions $i)$
and $ii)$ is satisfied), find $(u_2,v_2)$ such that
\vspace{-1mm}
\begin{eqnarray}
\hspace{-7mm}
(u_2,v_2)&\hspace{-1mm}=&\hspace{-1mm} 
{\mbox{arg min}\atop{u,v\thinspace \mbox{:} \thinspace u\ne u_1,v\ne v_1}} \thinspace \thinspace \thinspace
C\big(e_u^{(m)}(u,v)\big),
\end{eqnarray}

\vspace{-3mm}
and check for acceptance of the $(u_2,v_2)$ move. If this also cannot be
accepted, repeat the procedure for $(u_3,v_3)$, and so on. If all the 
$2kN$ moves are tabu, then all the 
$\textit{tabu\_matrix}$ entries are decremented by the minimum value in 
the $\textit{tabu\_matrix}$; this goes on till one of the moves becomes 
permissible. Let $(u',v')$ be the index of the neighbor with the minimum 
cost for which the move is permitted. The variables $q',q'',v''$ are 
implicitly defined by 
{\small $x_{u'}^{(m)}= a_{q'} = w_{v''}(x_{u'}^{(m+1)})$} (see
definition of $w_v(a)$ below Eqn.(ref{eq11})), and
$x_{u'}^{(m+1)}= a_{q''}$, where $a_{q'}, a_{q''} \in \mathbb{A}$. 

{\em Step 2:} 
After a move is done, the new solution vector is checked for repetition.
For the channel model in (\ref{SystemModelII}), repetition can be 
checked by comparing the ML costs of the solutions in the previous 
iterations. If there is a repetition, the length of the repetition from 
the previous occurrence is found, 

\newpage
the average length, $l_{rep}$, is 
updated, and the tabu period $P$ is modified as $P=P+1$.
If the number of iterations elapsed since the last change of the value of 
$P$ exceeds $\beta l_{rep}$, for a fixed $\beta > 0$, make $P=P-1$. The 
minimum value of $P$, however, will be 1. Note that this step, if executed, 
also qualifies as the one which changed $P$. After a move $(u',v')$ is 
accepted, if $\phi( \mathbf{x}^{(m+1)})< \phi(\mathbf{g}^{(m)})$, make
\vspace{-2mm}
\begin{eqnarray}
\textit{tabu\_matrix} \thinspace ((u'-1)M+q',v') \,\, = \,\, 0, \nonumber \\ 
\textit{tabu\_matrix} \thinspace ((u'-1)M+q'',v'') \,\, = \,\, 0, 
\end{eqnarray}
and $\mathbf{g}^{(m+1)} = \mathbf{x}^{(m+1)}$;
else,
\vspace{-2mm}
\begin{eqnarray}
\textit{tabu\_matrix}\thinspace ((u'-1)M+q',v') \,\, = \,\, P+1, \nonumber \\
\textit{tabu\_matrix}\thinspace ((u'-1)M+q'',v'') \,\, = \,\, P+1, 
\end{eqnarray}
and $\mathbf{g}^{(m+1)}=\mathbf{g}^{(m)}$.

{\em Step 3):} 
Update the entries of the $\textit{tabu\_matrix}$ as
\begin{eqnarray}
\hspace{-6mm}
\textit{tabu\_matrix}\thinspace(r,s) & \hspace{-1mm} = & \hspace{-1mm} \max \{\textit{tabu\_matrix}\thinspace(r,s)-1,0\},
\end{eqnarray}
for {\small $r=0,\cdots,2kM-1$}, {\small $s=0,\cdots,N-1.$}
$\mathbf{f}^{(m)}$ is updated as
\begin{eqnarray}
\mathbf{f}^{(m+1)} \,\,\, = \,\,\, \mathbf{f}^{(m)}+e_{u'}^{(m)}(u',v')\mathbf{R}_{u'},
\end{eqnarray} 
where $\mathbf{R}_{u'}$ is the ${u'}$th column of $\mathbf{R}$.

{\bfseries {\em Stopping criterion:}}
The algorithm can be stopped based on a fixed number of iterations.
Though convergence can be slow at low SNRs (typ. hundreds of iterations), 
it can be fast (typ. tens of iterations) at moderate to high SNRs. So 
rather than fixing a large number of iterations to stop the algorithm 
irrespective of the SNR, we use an efficient stopping criterion which 
makes use of the knowledge of the best ML cost in a given iteration,
as follows.

Since the ML criterion is to minimize ${\| \mathbf{Hx}-\mathbf{y}\|}^2$,
the minimum value of the objective function 
$\mathbf{x}^T\mathbf{H}^T\mathbf{Hx}-2\mathbf{x}^T\mathbf{H}^T\mathbf{y}$
is always greater than $-\mathbf{y}^T\mathbf{y}$. We stop the algorithm when the 
least ML cost achieved in an iteration is within certain range of the 
global minimum, which is $-\mathbf{y}^T\mathbf{y}$.
We stop the algorithm in the $m$th iteration, if the condition  
\begin{eqnarray}
\frac{|\phi(\mathbf{g}^{(m)})-(-\mathbf{y}^T\mathbf{y})|}{|-\mathbf{y}^T\mathbf{y}|} \,\,\, < \,\,\, \alpha _{1},
\end{eqnarray}
is met with at least $\textit{min\_iter}$ iterations being completed to
make sure the search algorithm has `settled.' The bound is gradually
relaxed as the number of iterations increase and the algorithm is
terminated when
\begin{eqnarray}
\frac{|\phi(\mathbf{g}^{(m)})-(-\mathbf{y}^T\mathbf{y})|}{|-\mathbf{y}^T\mathbf{y}|} \,\,\, < \,\,\, m\alpha _2.
\end{eqnarray}
In addition, we terminate the algorithm whenever the number of repetitions 
of solutions exceeds $\textit{max\_rep}$. Also, the maximum number of 
iterations is set to $\textit{max\_iter}$. We have found that use of the 
following stopping criterion parameters results in low complexity without 
compromising much on the performance (compared to a fixed number of 
iterations of 300) for 4-QAM: $\textit{min\_iter}=20$, 
$\textit{max\_iter}=300 $, $\textit{max\_rep}=75$, $\alpha _{1}=0.05$,
and $\alpha _{2}=0.0005$.

\vspace{-2.0mm}
\section{Simulation Results }
\label{sec4}
\vspace{-1.0mm}
In this section, we present the uncoded and coded BER performance of the 
RTS algorithm in decoding non-orthogonal STBCs
with $\delta=t=1$ 
(i.e., ILL) and $\delta=e^{\sqrt{5}{\bf j}}$, $t=e^{\bf j}$ (i.e., 
FD-ILL\footnote{Our simulation results show that the BER performance of 
FD-ILL and ILL STBCs with RTS decoding are almost the same.}).
The following RTS parameters are used in all the simulations: MMSE initial 
vector, {\small $P_0=2, \beta=1,0.1, \alpha_1=5\%, \alpha_2=0.05\%, 
\textit{max\_rep=75}, \textit{max\_iter}=300, \textit{min\_iter}=20$. }

\vspace{-4.0mm}
\subsection{Uncoded BER performance of RTS:}
\begin{figure}
\hspace{-6mm}
\includegraphics[width=0.55\textwidth,height=0.295\textheight]{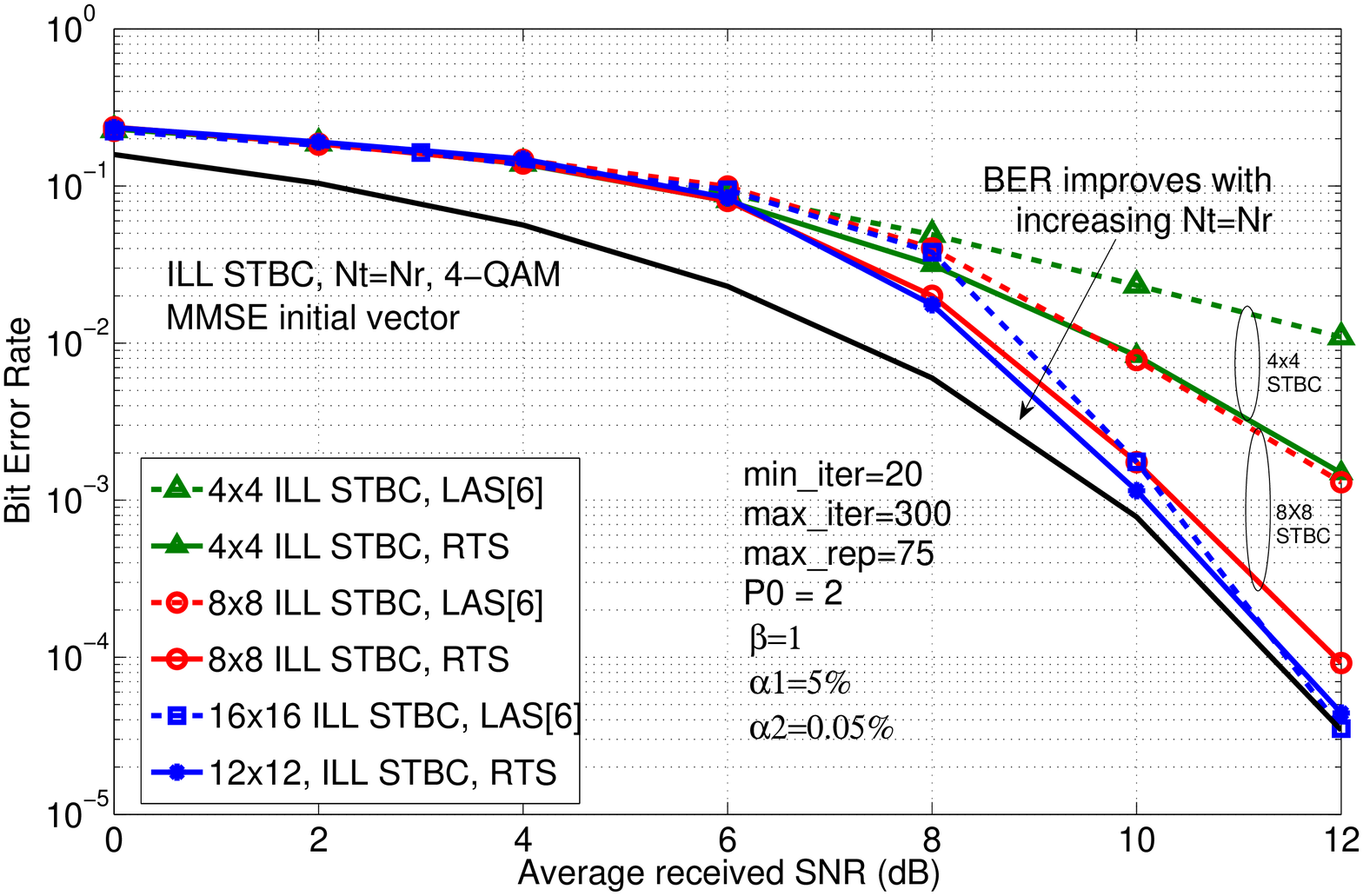}
\vspace{-7mm}
\caption{Uncoded BER of RTS decoding of $4\times 4$, $8\times 8$ and 
$12\times 12$ non-orthogonal STBCs from CDA. $N_t=N_r$, ILL STBCs 
($\delta=t=1$), 4-QAM. RTS parameters:
$P_0=2, \beta=1, \alpha_1=5\%, \alpha_2=0.05\%, \textit{max\_iter}=300,
\textit{min\_iter}=20$. {\em RTS achieves near SISO AWGN performance 
for increasing $N_t=N_r$ (i.e., STBC size). RTS performs better than LAS.}}
\vspace{-3mm}
\label{fig1}
\end{figure}

\vspace{-1.0mm}
{\bfseries {\em RTS versus LAS Performance:}}
In Fig. \ref{fig1}, we plot the uncoded BER of the RTS algorithm
as a function of average received SNR per receive antenna,
$\gamma$ \cite{jafarkhani}, in decoding $4\times 4$ (32 dimensions), 
$8\times 8$ (128 dimensions) and $12\times 12$ (288 dimensions) 
non-orthogonal ILL STBCs for 4-QAM and $N_t=N_r$. Perfect CSIR and 
i.i.d fading are assumed. For the same settings, performance of the 
LAS algorithm in \cite{jsac}-\cite{gcom08} are also plotted for
comparison. MMSE initial vector is used in both RTS and LAS.  
As a reference, we have plotted the BER performance on a 
SISO AWGN channel as well. From Fig. \ref{fig1}, the following 
interesting observations can be made:
\begin{itemize}
\item 	the BER of the RTS algorithm improves and approaches SISO AWGN 
	performance as $N_t\hspace{-0.5mm}=\hspace{-0.5mm}N_r$ (i.e., 
	STBC size) is increased; e.g., performance close to within 
	0.5 dB from SISO AWGN performance is achieved at $10^{-3}$ 
	uncoded {\small BER} in decoding $12\times 12$ STBC with 
	288 real dimensions.
\item	{\em RTS algorithm performs better than LAS algorithm} (see 
	RTS and LAS BER plots for $4\times 4$ and $8\times 8$ STBCs). 
	Further, while both RTS and LAS algorithms exhibit large system 
	behavior (i.e., BER improves as 
	$N_t\hspace{-0.5mm}=\hspace{-0.5mm}N_r$ is increased), RTS is 
	able to achieve nearness to SISO AWGN performance at $10^{-3}$ 
	BER with less number of dimensions than with LAS. This is evident 
	by observing that, while LAS requires 512 dimensions 
	($16\hspace{-0.10mm}\times \hspace{-0.10mm}16$ STBC) to achieve 
	1 dB closeness to SISO AWGN performance at $10^{-3}$ BER, RTS is 
	able to achieve even 0.5 dB closeness with just 288 dimensions 
	($12\hspace{-0.5mm}\times \hspace{-0.5mm}12$ STBC). RTS is able
	to achieve this better performance because, while the 
	bit/symbol-flipping strategies are similar in both RTS and LAS, 
	the inherent escape strategy in RTS allows it to move out of local
	minimas and move towards better solutions. Consequently, RTS incurs 
	some extra complexity compared to LAS, without increase in the 
	order of complexity. 
\end{itemize}

\begin{figure}
\hspace{-6mm}
\includegraphics[width=0.55\textwidth,height=0.295\textheight]{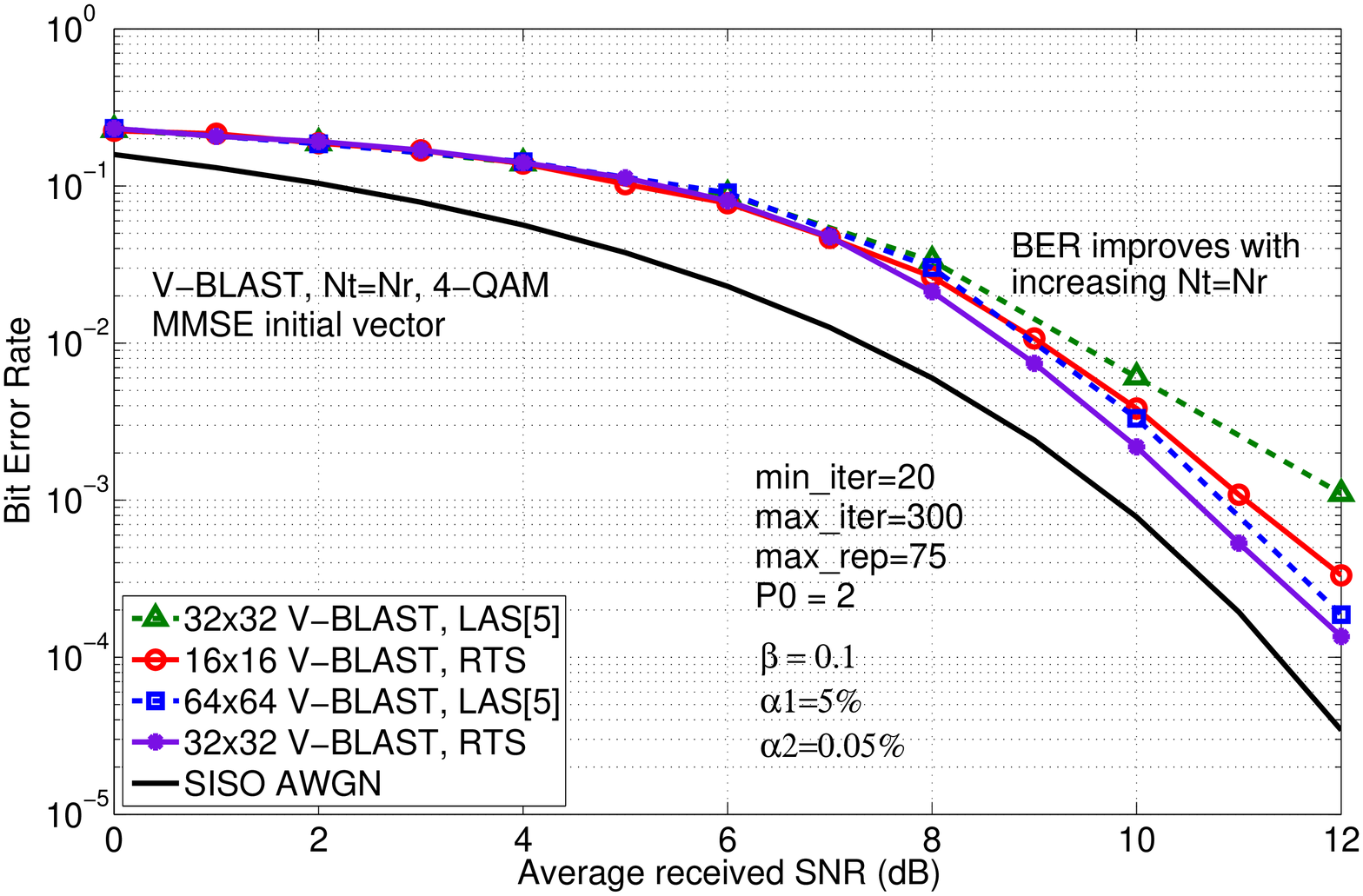}
\vspace{-7mm}
\caption{Uncoded BER of RTS detection of V-BLAST with $N_t=N_r$ and 4-QAM. 
RTS parameters: $P_0=2, \beta=0.1, \alpha_1=5\%, \alpha_2=0.05\%, 
\textit{max\_iter}=300, \textit{min\_iter}=20$. {\em RTS achieves near 
SISO AWGN performance for increasing $N_t=N_r$. RTS performs better 
than LAS.}}
\vspace{-3mm}
\label{fig2}
\end{figure}

{\em RTS performance in V-BLAST:}	
A similar observation can be made with uncoded BER of RTS detection in 
V-BLAST in Fig. \ref{fig2} for $N_t\hspace{-0.0mm}= \hspace{0.0mm}N_r$ 
and 4-QAM. From Fig. \ref{fig2}, it is seen that LAS requires 128 
dimensions ($64\hspace{-0.25mm}\times \hspace{-0.25mm} 64$ V-BLAST) to 
achieve performance within 1 dB of SISO AWGN performance at $10^{-3}$ 
BER, whereas RTS is able to achieve even better closeness with just 64 
dimensions ($32\hspace{-0.0mm}\times \hspace{-0.0mm} 32$ V-BLAST). In 
summary, the ability to achieve near SISO AWGN performance at less 
dimensions than LAS is an attractive feature of RTS. 

\vspace{-3mm}
\subsection{Turbo coded BER performance of RTS} 
\vspace{-1mm}
Figure \ref{fig3} shows the rate-3/4 turbo coded BER of RTS decoding 
of $12\times 12$ non-orthogonal ILL STBC with $N_t=N_r$ and 4-QAM
(corresponding to a spectral efficiency of 18 bps/Hz), under 
perfect CSIR and i.i.d fading. The theoretical minimum SNR required 
to achieve 18 bps/Hz spectral efficiency on a 
$N_t\hspace{-0.25mm}=\hspace{-0.25mm}N_r\hspace{-0.25mm}=\hspace{-0.25mm}12$
MIMO channel with perfect CSIR and i.i.d fading is 4.27 dB (obtained
through simulation of the ergodic capacity formula \cite{jafarkhani}).
>From Fig. \ref{fig3}, it is seen that RTS decoding is able to achieve 
vertical fall in coded BER close to within about 5 dB from the theoretical 
minimum SNR, which is good nearness to capacity performance. This nearness
to capacity can be further improved by 1 to 1.5 dB if soft decision 
values, proposed in \cite{isit08}, are fed to the turbo decoder.

\vspace{-3.0mm}
\subsection{Iterative RTS Decoding/Channel Estimation}
\begin{figure}
\hspace{-7mm}
\includegraphics[width=0.55\textwidth,height=0.295\textheight]{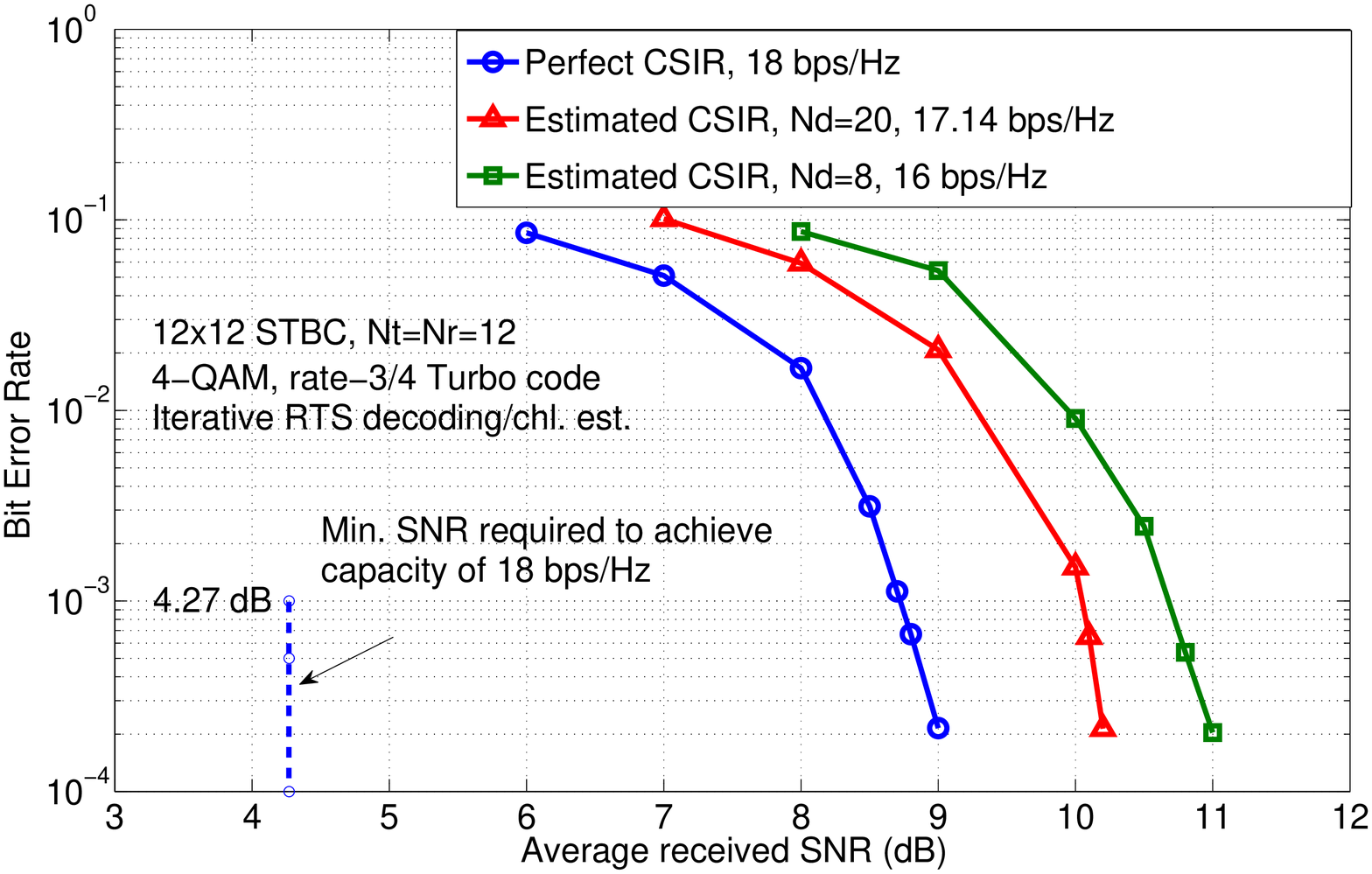}
\vspace{-7mm}
\caption{Turbo coded BER of RTS decoding of $12\times 12$ non-orthogonal
ILL STBC with $N_t=N_r$, 4-QAM, rate-3/4 turbo code, and 18 bps/Hz.
RTS parameters: $P_0=2, \beta=1, \alpha_1=5\%, \alpha_2=0.05\%, 
\textit{max\_iter}=300, \textit{min\_iter}=20$. {\em BER of RTS with 
estimated CSIR approaches close to that with perfect CSIR for increasing 
$N_d$ (i.e., slow fading).}}
\vspace{-0mm}
\label{fig3}
\end{figure}

\begin{figure}
\vspace{1mm}
\centering
\includegraphics[width=3.4in]{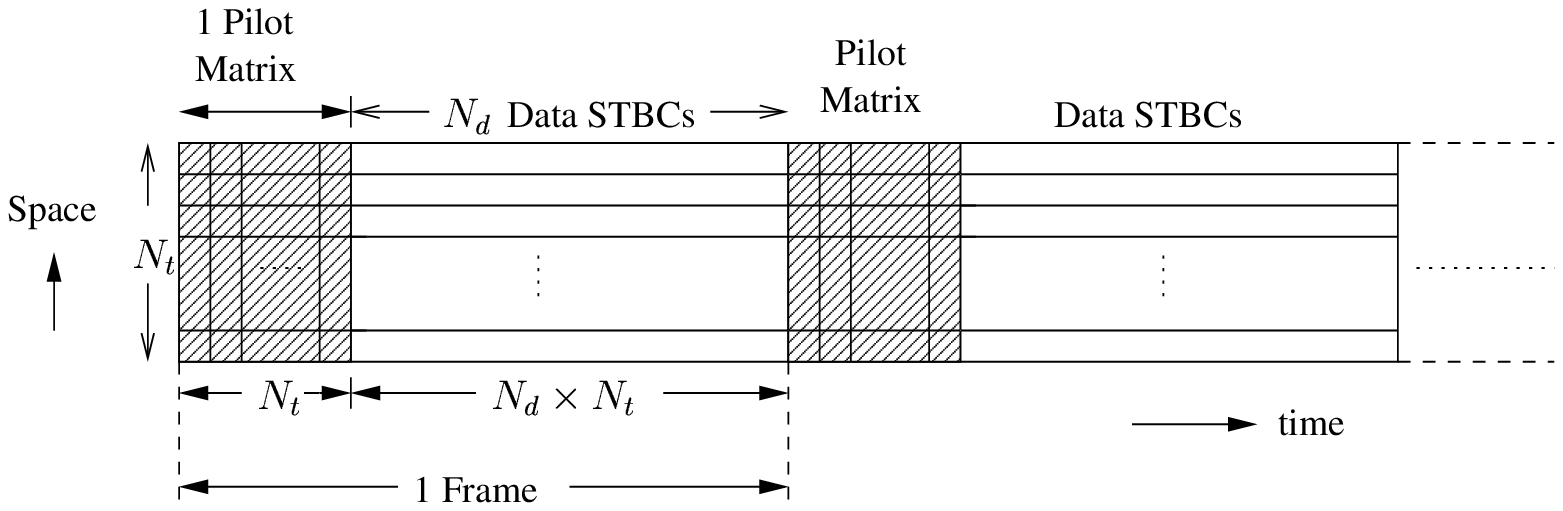}
\vspace{-5mm}
\caption{Transmission scheme with one pilot matrix followed by
$N_d$ data STBC matrices in each frame. }
\label{fig4}
\vspace{-2.5mm}
\end{figure}

\vspace{-0.5mm}
Next, we relax the perfect CSIR assumption by considering a training 
based iterative RTS decoding/channel estimation scheme. Transmission is
carried out in frames, where one {\small $N_t\times N_t$} pilot matrix
(for training purposes) followed by $N_d$ data STBC matrices are sent
in each frame as shown in Fig. \ref{fig4}. One frame length, $T$,
(taken to be the channel coherence time) is $T=(N_d+1)N_t$ channel
uses. The proposed scheme works as follows \cite{zaki}: $i)$ obtain an 
MMSE estimate of the channel matrix during the pilot phase, $ii)$ use the
estimated channel matrix to decode the data STBC matrices using RTS 
algorithm, and $iii)$ iterate between channel estimation and RTS
decoding for a certain number of times. For $12\times 12$ ILL STBC, 
in addition to perfect CSIR performance, Fig. \ref{fig3} also shows 
the performance with CSIR estimated using the above iterative RTS 
decoding/channel estimation scheme for $N_d=8$ and $N_d=20$. 
2 iterations between RTS decoding and channel estimation are used. With
$N_d=20$ (which corresponds to large coherence times, i.e., slow fading)
the BER and bps/Hz with estimated CSIR get closer to those with
perfect CSIR.

\vspace{-3.5mm}
\subsection{Effect of MIMO Spatial Correlation}
\begin{figure}
\centering
\includegraphics[width=0.525\textwidth,height=0.295\textheight]{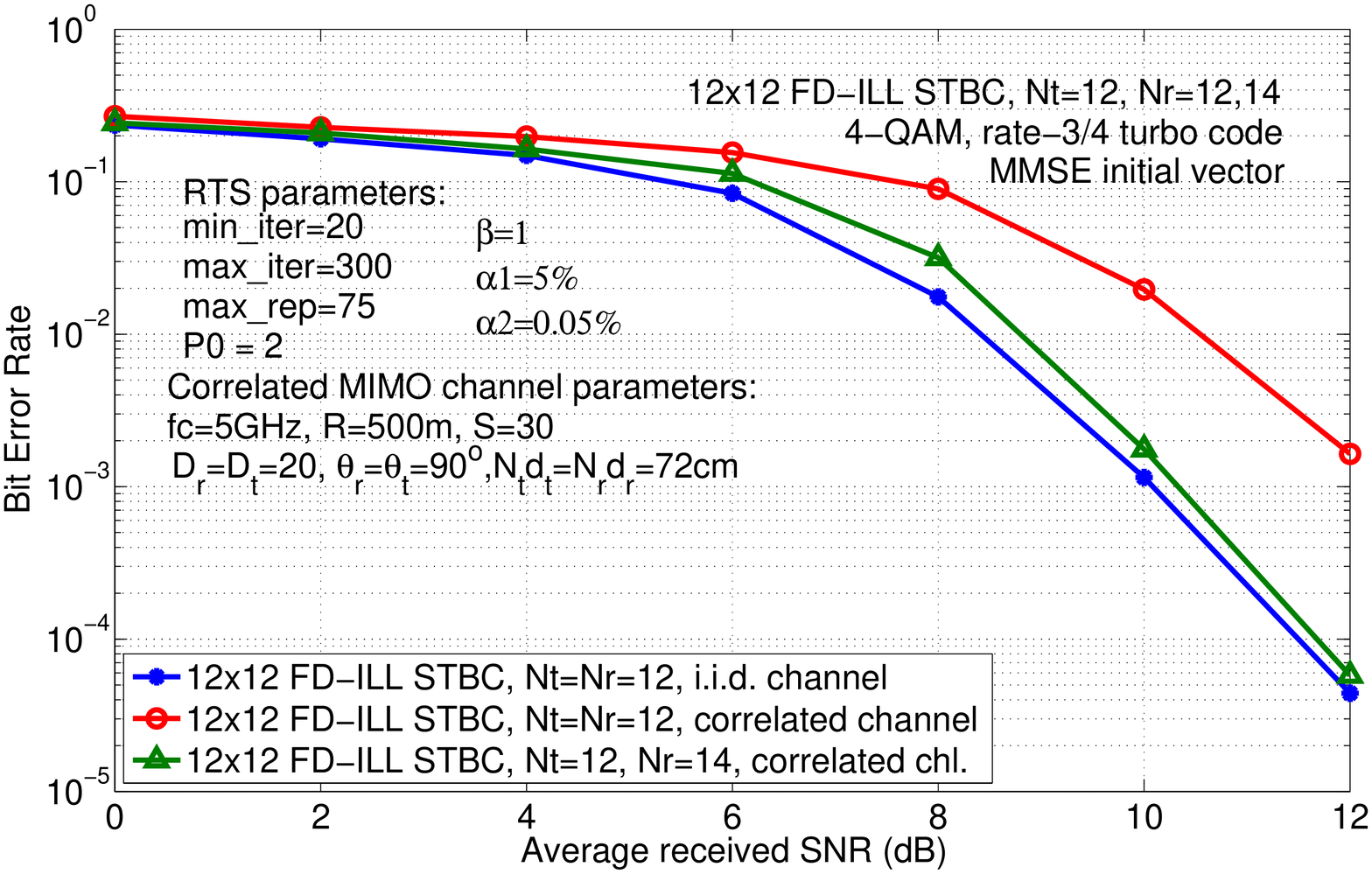}
\vspace{-7mm}
\caption{
Effect of spatial correlation on the performance of RTS decoding of
$12\times 12$ FD-ILL STBC with $N_t=12$, $N_r=12,14$, 4-QAM,
rate-3/4 turbo code, 18 bps/Hz. Correlated MIMO channel
parameters: $f_c=5$ GHz, $R=500$ m, $S=30$, $D_t=D_r=20$ m,
$\theta_t=\theta_r=90^\circ$, $N_rd_r=N_td_t=72$ cm.
{\em Spatial correlation degrades achieved diversity order compared to 
i.i.d. Increasing $N_r$ alleviates this performance loss.} }
\vspace{-5mm}
\label{fig5}
\end{figure}

\vspace{-1mm}
In Figs. \ref{fig1} to \ref{fig3}, we assumed i.i.d fading. But spatial 
correlation at transmit/receive antennas and the structure of scattering 
and propagation environment can affect the rank structure of the MIMO 
channel resulting in degraded performance \cite{mimo1},\cite{mimo2}. We 
relaxed the i.i.d. fading assumption by considering the correlated MIMO 
channel model proposed by Gesbert 
et al in \cite{mimo2}, which takes into account carrier frequency 
($f_c$), spacing between antenna elements {\small ($d_t,d_r$)}, distance 
between transmit and receive antennas ($R$), and scattering environment. 
In Fig. \ref{fig5}, we plot the uncoded BER of RTS decoding of $12\times 12$ 
FD-ILL STBC with perfect CSIR in $i)$ i.i.d. fading, and $ii)$ correlated 
MIMO fading model in \cite{mimo2}. It is seen that, compared to i.i.d 
fading, there is a loss in diversity order in spatial correlation for 
$N_t=N_r=12$; further, use of more receive antennas ($N_r=14, N_t=12$), 
without increase in the receiver aperture,
alleviates this loss in performance. Finally, we note that have carried 
out simulations of RTS decoding for 16-QAM as well, where similar results 
reported here for 4-QAM are observed. The RTS decoding can be used to 
decode {\em perfect codes} \cite{perf06},\cite{perf07} of large dimensions 
as well. 

\vspace{-4mm}
\section{Conclusions}
\label{sec5}
\vspace{-1mm}
We presented a reactive tabu search based low-complexity algorithm for 
decoding high-rate non-orthogonal STBCs having large dimensions that 
can achieve high spectral efficiencies of the order of tens of bps/Hz. 
The RTS algorithm was shown to achieve near SISO AWGN uncoded BER 
performance as well as near-capacity turbo coded BER performance in
non-orthogonal STBC MIMO systems with large dimensions. The algorithm 
performed well with estimated CSIR using a training-based iterative 
decoding/channel estimation scheme. In addition, the algorithm could
perform well in the presence of MIMO spatial correlation when more
receive dimensions are used. Comparing the performance of RTS algorithm
with LAS algorithm (which we presented recently), we pointed out that
the ability to achieve near SISO AWGN performance at less
dimensions than LAS is an attractive feature of RTS.

\end{document}